\documentclass{svproc}
\usepackage{url}
\usepackage{amsmath}
\usepackage{amssymb}
\usepackage{amsfonts}
\usepackage{mathtools}
\usepackage{dsfont}
\usepackage{adjustbox}
\usepackage{graphicx}

\begin{document}

\mainmatter              
\title{Competitiveness of Formula 1 championship from 2012 to 2022 as measured by Kendall corrected evolutive coefficient}
\titlerunning{Competitiveness of Formula 1 championship from 2012 to 2022}  

%
\author{Francisco Pedroche\inst{1}}
\authorrunning{Francisco Pedroche} 
%
\tocauthor{Francisco Pedroche}
\institute{Institut Universitari de Matem\`{a}tica Multidisciplin\`{a}ria, \\ Universitat Polit\`{e}cnica de Val\`{e}ncia\\
 Cam\'{\i} de Vera s/n, 46022, Val\`{e}ncia, Spain\\
\email{pedroche@imm.upv.es},\\ WWW home page:
\texttt{http://personales.upv.es/pedroche/}}

\maketitle              

\begin{abstract}
In this paper we analyze the FIA formula one world championships from 2012 to 2022 taking into account
the drivers classifications and the constructors ({\em teams}) classifications of each Grand Prix.
The needed data consisted of 22 matrices of sizes ranging from $25 \times 20$
to $10 \times 19$
that have been elaborated from the GP classifications extracted from the official FIA site.
We have used
the  Kendall corrected evolutive coefficient, recently  introduced, as a measure of Competitive Balance (CB) to study the evolution of the competitiveness along the years in both
drivers and teams championships. In addition,
we have compared the CB of F1 championships and two major European football leagues from the seasons 2012-2013 to 2022-2023.
\end{abstract}
\section{Introduction}

A  \emph{ranking} naturally appears when we sort elements, being this a key action in more activities such as analysis of sport competitions \cite{Basi}, economic time series \cite{KeHi}, comparison of algorithms performance
\cite{Vecek}, etc.
Series of rankings can be studied from different perspectives. For example,  to analyse sorting algorithms \cite{Knuth},
to define measures of disarray \cite{Diacon},
to use rank transformation to develop nonparametric methods in Statistics \cite{Conover}, to {\em learn to rank} in machine learning applications \cite{Cheng}, etc.
In this paper we are interesting in characterising a series of rankings by giving a coefficient that measures the disarray along the series in the classic manner of
\cite{Kendall38}. Specifically, we follow the definitions of \cite{Pe20}, \cite{PeCrGaRoSa} and \cite{Criado}.

\section{Kendall corrected evolutive coefficient}

The Kendall corrected evolutive coefficient, denoted by $\widehat{\tau}_{ev}^{\bullet}$,  was introduced in \cite{Pe20}.
It takes as input a series of $m$ rankings (with at most $n$ elements) that can be\emph{ complete } (that is, the $n$ elements are ranked in all the rankings) or incomplete.
In addition, we consider the existence or not
of ties between the ranked elements.  Kendall corrected evolutive coefficient can be considered as an extension of a correlation coefficient of two rankings applied to $m$ rankings and therefore, as output, $\widehat{\tau}_{ev}^{\bullet}$ gives a real number in $ [-1,1]$.

The coefficient $\widehat{\tau}_{ev}^{\bullet}$  reduces to some particular coefficients that are well documented and can be found in the literature. For example,
when $m=2$ and the rankings are complete and with no ties, then $\widehat{\tau}_{ev}^{\bullet}$ reduces to the classical Kendall's $\tau$ coefficient of
disagreement (see \cite{Kendall38},
\cite{Kendall70},
\cite{KeBa39}) that can be written as
\begin{equation}\label{eqtauPQ}
\tau = \frac{2(P-Q)}{n (n-1)}
\end{equation}
where $P$ is the number of pair of elements that keep its relative order from the first ranking to
the second one and $Q$ is the number of pairs of elements that change its order.
For example, taking $n=3$, the rankings  ${\bf a}=[1,2,3]$ and ${\bf b}=[3,2,1]$ have an associated $\tau=-1$ and the rankings
${\bf a}=[1,2,3]$ and  ${\bf b}=[1,2,3]$ have an associated $\tau=1$. When $m=2$ and the rankings are complete and with ties, then  $\widehat{\tau}_{ev}^{\bullet}$ is related to the {\em Kendall distance with penalty parameter $p \in [0,\frac12]$} defined in \cite{Fa06}.
When $m>2$ and the rankings are complete and with ties, then  $\widehat{\tau}_{ev}^{\bullet}$ reduces the {\em corrected
evolutive Kendall distance with penalty parameter $p$}  introduced in \cite{PeCrGaRoSa}.

In sport competitions it is most used the term {\em Competitive Balance} (CB) to measure the balance between the teams \cite{Zimba},
\cite{Owen}. A high measure of CB means that the competition is highly interesting since it is very difficult to predict the result of a match (or a race, in our case), while a low measure of CB means that the competition is very predictable, and therefore {\em boring} (see. \cite{Mastro}, \cite{Manasis}, \cite{Garc}, \cite{Judde}, \cite{Basi}, \cite{Budz}).
In this regard it is more convenient to use the measure called {\em Normalized Strength} (borrowed from complex networks terminology, see \cite{Criado}, \cite{Barrat}), and that we define here by
\begin{equation}\label{eqNS}
NS = \frac{1 - \widehat{\tau}_{ev}^{\bullet}}{2}
\end{equation}
Note that $NS$ is a normalized index, $NS \in [0,1]$, and its value can be considered as a measure of CB. We will use this index in our analysis.
The interested reader may find the precise definition of $\widehat{\tau}_{ev}^{\bullet}$ in \cite{Pe20} but we omit the details for the sake of brevity.

\section{Formula One World Championships}

Formula One (also known as Formula 1 or F1) organised by the F\'ed\'eration Internationale de l'Automobile (FIA)
is a well-known international racing for cars \cite{link-fia}. The drivers championship began in the season of 1950, while the constructors championship began in 1958.
Along the years, there has been some
modifications both in the format and in the rules that the participants must accomplish.

A Formula One season consists of a series of races, each of them known as Grand Prix (denoted as GP), that take place in several countries.
For example, the F1 2022 season consisted of 22 GP and participated 10 teams and 22 drivers. A GP is held on a weekend.
On friday and saturday some qualifying sessions fix the starting order (\emph{the grid}) for the GP race that occurs on Sunday.
In this paper we are interested only in the ranking corresponding to this GP races.
This ranking is decided based on the timing of
each driver and he receives a quantity of points depending on his ranking.
From 2010 to 2018 the sharing of the points was given as shown in Table \ref{ta:points}.
The points assigned to the constructors in a GP is the sum of the points of the two drivers of the team that participated in that GP.
\begin{table}[!ht]
\vspace*{0.1cm}
\caption{Points scoring sharing since 2010}
\begin{center}
\begin{tabular}{|c|c|c|c|c|c|c|c|c|c|c|}
\hline
1st	& 2nd	& 3rd	& 4th&	5th&	6th&	7th&	8th&	9th&	10th\\
\hline
25	& 18	& 15& 	12& 	10& 	8& 	6& 	4& 	2& 	1\\
\hline
\end{tabular}
\label{ta:points}
\end{center}
\end{table}

From 2019 one additional point is given to the pilot that occupied a position in the top ten and furthermore has the fastest lap in the race.
FIA has some rules to break ties between the pilots and therefore the ranking of the drivers can be considered as ranking with no ties.
Note, therefore that each GP has its own classification.
The final ranking (that is, the F1 Championship) of the season is made by accumulating the points of each GP,
and, again, some rules are applied to break the ties, if any. Our collection of rankings are precisely the rankings of each GP in a season, both for drivers and constructors. We use these series of rankings to compute
the corresponding $\widehat{\tau}_{ev}^{\bullet}$ of that season, and then the corresponding $NS$. We precisely describe the used rankings in the next section.

\section{Description of the rankings}

\begin{table}[!ht]
\caption{Number of drivers, teams and GP in each analyzed F1 Championship}
\begin{center}
\begin{tabular}{|l|c|c|c|c|c|c|c|c|c|c|c|}
\hline
Year & 2012 & 2013 & 2014 & 2015 & 2016 & 2017& 2018 & 2019& 2020& 2021 & 2022  \\
\hline
Drivers & 25 & 23 & 24 & 22 & 24 & 25 & 20 & 20 & 23 & 21 & 22 \\
\hline
Teams & 12 & 11& 11& 10 & 11& 10 & 11& 10& 10 & 10& 10 \\
\hline
GP's & 20 & 19& 19& 19& 21& 20&21& 21& 17& 22& 22 \\
\hline
\end{tabular}
\label{ta:nGP2}
\end{center}
\end{table}

We have selected the F1 classifications from 2012 to 2022.
Our criterium to select our dataset is based on taking the GP classifications of championships in where 1) the regulations does not vary too much,
2) the distribution of points (e.g. as given by Table \ref{ta:points}) is quite stable,
3) the number of GP does not vary too much and 4) that the standings can be easily retrieved from the official FIA site \cite{link-fia}.
For example, the 2012 season can be retrieved from the FIA site \cite{season2012}.
In  Table \ref{ta:nGP2}  we show the number of drivers in each championship jointly with the number of GP in that year.

To describe our rankings we use the following notation (see  \cite{Yoo}, \cite{Pe20}).
Let $V =\{ v_1, v_2, \cdots, v_n \}$ be the objects to be ranked, with~$n>1$. The~ranking is given by
\begin{equation}\label{def:a}
{\bf a} = [a_1,a_2, \cdots, a_n]
\end{equation}
where $a_i$ is the position of $v_i$ in the ranking. Note that if $a_i=a_j$, then $v_i$ and $v_j$ are tied. If~$v_i$ is not ranked, then it is denoted as $a_i=\bullet$.

\begin{table}[!ht]
\caption{Drivers' name, nationality, and ${\bf a}_i$ vector for three of the first GP of FIA 2012 World Championship. Elaborated from \cite{link-fia}. Note that $\bullet$ means that the driver did not start or did not finish the race. The rankings are \emph{incomplete rankings with no ties}. The order of the drivers in the first column follows the (final) classification of the constructors championship.
The drivers Raikkonen,  Grosjean and  D'Ambrosio belong to the same team (Lotus F1) while the rest of teams contributed with two drivers in the whole GP rankings of this championship.}
\begin{center}
\begin{tabular}{|l|c|c|c|c|}
\hline
Driver	& Nat &	GP1& 	GP2& 	GP3\\
\hline
Sebastien Vettel &	DEU & 	2	& 11 & 	5\\
Fernando Alonso	&ESP	&5&	1 &	9\\
Kimi Raikkonen	&FIN	&7&	5&	14\\
Lewis Hamilton	&GBR	&3&	3&	3\\
Jenson Button	&GBR	&1&	14&	2\\
Mark Webber	&AUS	&4&	4&	4\\
Felipe Massa	&	BRA	 & $\bullet$ &	15&	13\\
Romain Grosjean		&FRA	&$\bullet$&	$\bullet$&	6\\
Nico Rosberg		&DEU	&12	&13&	1\\
Sergio Perez		&MEX	&8&	2 &	11\\
Nico Hulkenberg		&DEU	&$\bullet$&	9&	15\\
Kamui Kobayashi		&JPN	&6&	$\bullet$&	10\\
Michael Schumacher		&DEU&	$\bullet$&	10&	$\bullet$\\
Paul Di Resta		&GBR	&10&	7&	12\\
Pastor Maldonado		&VEN&	13&	19&	8\\
Bruno Senna		&BRA&	16&	6&	7\\
Jean-Eric Vergne	&FRA	&11&	8&	16\\
Daniel Ricciardo&	AUS	&9&	12	&17\\
Vitaly Petrov	&RUS&	$\bullet$&	16&	18\\
Timo Glock	&DEU&	14&	17&	19\\
Charles Pic	&FRA&	15&	20&	20\\
Heikki Kovalainen&	FIN	&$\bullet$&	18&	23\\
J\'{e}r\^{o}me D'Ambrosio&	BEL	&$\bullet$&	$\bullet$&	$\bullet$\\
Narain Karthikeyan&	IND	&$\bullet$&	 22& 	22\\
Pedro De la Rosa	&ESP&	$\bullet$	& 21	& 21\\
\hline
\end{tabular}
\label{ta:losai}
\end{center}
\end{table}

\subsection{Drivers ranking}

From the FIA site, we can retrieve the drivers classification for each GP of the considered championship.
In these classifications we can see the ranking, the points obtained by each driver, and a note indicating whether the driver has finished
the race or not. To construct our drivers ranking we consider that a driver that has not finished the race (or has not even start it) is an absent element in our ranking, and therefore it is indicated by $\bullet$. For example, in Table \ref{ta:losai} we show our notation to describe
the first three rankings of the 2012 championship.

\subsection{Constructors ranking}

From the FIA site we can retrieve the constructors classification for each GP of the considered championship.
The points given to a constructor consist of the sum of the points of the two drivers of the corresponding team in each GP.
In this case the FIA site offers the points obtained by each constructor. This gives us the opportunity to create two types of rankings, being the interest to see how our measure $NS$ is affected by these types. The two considered methods are the following:

{\bf Method 1}: We consider that the constructors that have $0$ points are tied in the last position.

{\bf Method 2}: We consider that the constructors that have $0$ points are absent elements.

As an example, in Table \ref{ta:constpuntos} we show the constructors name, scoring and ${\bf a}_i$
vectors (by using Method 1 and Method 2) for the first three GP of FIA 2012 World Championship.


\begin{table}[!ht]
\caption{Constructor's name, scoring and ${\bf a}_i$ vectors (by using Method 1 and Method 2) for the first three GP of FIA 2012 World Championship. The order of the teams in the first column follows the (final) classification of the championship.}
\begin{center}
\begin{tabular}{|l|c|c|c|c|c|c|c|c|c|}
\hline
 & \multicolumn{3}{|c|}{Score } & \multicolumn{3}{|c|}{ Method 1 } &  \multicolumn{3}{|c|}{ Method 2 }  \\
\cline{2-10}
Constructors &  GP1 & GP2 & GP3  &  GP1 & GP2 & GP3 &  GP1 & GP2 & GP3 \\
\hline
Red Bull Racing&              30&12 &22  &   2&4 &3 & 2&4 &3 \\
Scuderia Ferrari&             10&25 &2 & 4&1 &6 &4&1 &6 \\
Vodafone McLaren Mercedes&    40&15 & 33 &  1&3 & 1  &  1&3 & 1\\
Lotus F1 Team&                6&10 &8&  5&5 &5 &  5&5 &5\\
Mercedes AMG Petronas F1 Team&0&1 & 25 & 8&8 & 2  & $\bullet$&8 & 2\\
Sauber F1 Team&               12&18 &1 &  3&2 &7  &  3&2 &7\\
Sahara Force India F1 Team&   1&8 &0 &  7&6 &8 &  7&6 &$\bullet$\\
Williams F1 Team&             0&8 &10 &   8&6 &4 &   $\bullet$&6 &4\\
Scuderia Toro Rosso&          2&4 &0  &   6&7 &8&   6&7 &$\bullet$\\
Caterham F1 Team&             0&0&0 & 8&8&8& $\bullet$&$\bullet$&$\bullet$\\
Marussia F1 Team&             0&0&0 &  8&8&8& $\bullet$&$\bullet$&$\bullet$\\
HRT F1 Team&                  0&0&0  &  8&8&8& $\bullet$&$\bullet$&$\bullet$\\
\hline
\end{tabular}
\label{ta:constpuntos}
\end{center}
\end{table}

\section{Results}

\subsection{Comparison of constructors and drivers championships}\label{comp}

In order to compare the\emph{ competitivity balance} of the GP of drivers and constructors  we have computed $NS$, given by (\ref{eqNS}) for the GP standings from 2012 to 2022 for drivers and for constructors (with Method 1 and Method 2).  The results are shown in Table \ref{ta:NS}.


\begin{table}[!ht]
\caption{$NS$ for the series of GP of the Championships from 2012 to 2022 for drivers and constructors.}
\begin{center}
\begin{tabular}{|c|c|c|c|}
\hline
Year  &  $NS$  Drivers     & $NS$ Constructors   & $NS$ Constructors  \\
  \hspace*{1cm}     &   & Method 1          & Method 2  \\
\hline
2012& 0.2561  & 0.2456  &   0.4052\\
2013& 0.2136  &  0.1924 &  0.3421\\
2014& 0.1913  &  0.1616& 0.3106\\
2015& 0.2270  & 0.2722&  0.2350\\
2016& 0.2065  & 0.2218&  0.2143\\
2017& 0.2140  & 0.2632&  0.2179 \\
2018& 0.2188  & 0.2559&  0.1886 \\
2019& 0.2157  & 0.2772&  0.2596 \\
2020& 0.2436  & 0.2562&   0.3652 \\
2021& 0.2270  & 0.2413&  0.2715 \\
2022& 0.2092  & 0.2455&   0.2376\\
\hline
\end{tabular}
\label{ta:NS}
\end{center}
\end{table}

\begin{figure}[h]
\includegraphics[width=\textwidth]{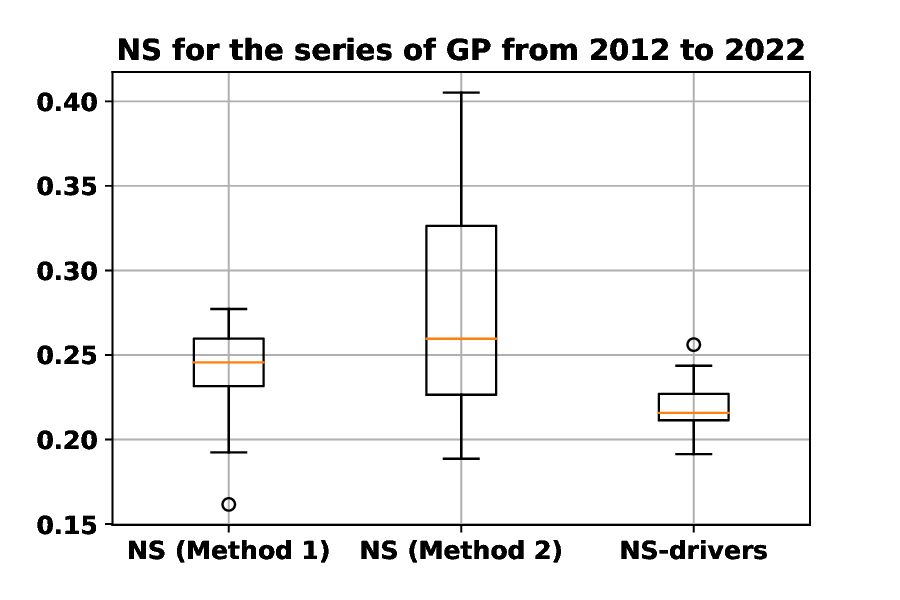}
\caption{Box-and-whiskers diagram for $NS$ for the series of GP of the championships
from 2012 to 2022 for drivers and constructors (by using the two methods explained on the text).}
\label{fig-box}
\end{figure}

The data on Table \ref{ta:NS} can be resumed on the box-and-whiskers plot shown on Figure  \ref{fig-box}. In more detail,
the mean values of $NS$ on the period 2012-2022, and the corresponding sample standard deviation, $s$, are as follows:

\noindent
Mean value of $NS$ for drivers:  $0.2203$, ($s= 0.018$).

\noindent
Mean value of $NS$ for constructors (Method 1):  $0.2394$, ($s= 0.035$).

\noindent
Mean value of $NS$ for constructors (Method 2):    $0.2771$, ($s= 0.070$).

Let us consider that $NS$ is a random variable. By computing the Shapiro-Wilk test for normality \cite{Shap}
we obtain the p-values $0.61$, $0.08$ and $0.44$ for the corresponding $NS$ series for drivers, and constructors (Method 1 and Method 2) respectively.
Therefore we cannot reject the normality of the distribution of $NS$ of the corresponding samples.
Regarding the mean values of $NS$ for constructors by using Method 1 and Method 2, since they come from the same data (as an example, the scores in Table \ref{ta:constpuntos} give us the corresponding values for Method 1 and Method 2) we can use a comparison method for means coming from paired data. By using a t-test we obtain a p-value of $0.18$ and therefore we cannot reject that the means are equal with a confidence interval of $95\%$. Since the value of the variances does not have a ratio major than 4 we can use the t-test for comparing the mean of $NS$ by using Method 1, and the corresponding $NS$ for drivers.
We obtain that the p-value is 0.12 and therefore we cannot reject the null hypothesis that the means are equal.
All in all we have the statistically the three values of $NS$ are not different, with a confidence interval of 95\%.



%

%

\subsection{Comparison of competitiveness between F1 championships and two major European football leagues}

A competitive balance measure like $NS$, based on sport ranking series, can be used to compare the CB of two different sports.
For example, by computing the coefficient $NS$ for two major European football leagues (Spanish League -commercially known as Laliga Santander in the season 2022/23-, and
the English Premier  league) we obtain the results shown in Table \ref{ta:NSFut}. We have used the series of standings from the season 2012-2013 to the season 2022-2023
for both the Spanish League (retrieving the data from the links on \cite{wiki-esp})
and Premier League (retrieving the data from \cite{cero-premier}). The summary for the football leagues in the studied period is the following:

\noindent
Mean value of $NS$ for Spanish league:  $0.059$, ($s= 0.0094$).

\noindent
Mean value of $NS$ for Premier league:  $0.056$, ($s= 0.0062$).

As a consequence, by using the results on section \ref{comp} for $NS$ of drivers and $NS$ of constructors by using Method 1,
we obtain that the mean value of $NS$ for the F1 championships is about four times greater than the values of $NS$ corresponding to the
analyzed football leagues.

\begin{table}[!ht]
\caption{$NS$ values for two European football leagues from season 2012/2013 to season 2022/2023.}
\begin{center}
\begin{tabular}{|c|c|c|}
\hline
Year  &  $NS$      & $NS$    \\
\hspace*{1cm}      &  Spanish league& Premier league  \\
\hline
2012&  0.0615 &0.0514\\
2013& 0.0593 &0.0656\\
2014& 0.0546 &0.0597\\
2015& 0.0613 &0.0563\\
2016& 0.0435 &0.0589\\
2017& 0.0589 &0.0550\\
2018& 0.0688&0.0489\\
2019& 0.0600&0.0643\\
2020& 0.0757&0.0583\\
2021& 0.0440&0.0461\\
2022& 0.0595 & 0.0515\\
\hline
\end{tabular}
\label{ta:NSFut}
\end{center}
\end{table}

\section{Conclusions}

In this communication we have shown how to apply a recently introduced metric to calculate a measure of the competitive balance (CB) associated to Formula 1 championships, by taking
into account the standings of the Grand Prix that compose each championship.
We have introduced to methods (called Method 1 and Method 2) to compute the CB values of the F1 Constructors Championship in the period 2012-2022.
We have obtained that these two
methods do not offer mean values that can be considered statistically different.
We think that this shows a good behaviour of our metric since both Method 1 and Method 2
are obtained by computing a linear combination from the same set of data (the F1 Drivers Championship) but with different treatment of the constructors that finish with zero points
in a Grand Prix. We also have obtained that the CB of the F1 Drivers Championship and F1 Constructors Championship show similar values on the studied period, but with a slightly higher mean value for the Constructors Championship. As an example of the power of our metric, we have compared the CB of two different sports: the
Formula 1 championships from 2012 to 2022 and the Spanish football league and Premier football league on the seasons 2012-2013 to 2022-2023.
Our results show that the mean value of CB for the F1 championships is about four times greater than the values of CB corresponding to the
analyzed football leagues.

%
%


\begin{thebibliography}{6}


%
%
%
%
%


\bibitem{Barrat}
Barrat, A., Barthélemy, M.,   Pastor-Satorras, R., Vespignani, A.:
The architecture of complex weighted networks,
Proceedings of the National Academy of Sciences, 101(11), 3747-3752 (2004).
\url{doi:10.1073/pnas.0400087101}



\bibitem{Basi}
Basini, F.,
Tsouli, V.,
Ntzoufras, I.,
Friel, N.: Assessing competitive balance in the English Premier League for over forty seasons using a stochastic block model, Journal of the Royal Statistical Society Series A: Statistics in Society, qnad007 (2023). \url{doi:10.1093/jrsssa/qnad007}

\bibitem{Budz}
Budzinski, O.,  Feddersen, A.: Measuring Competitive Balance in Formula One Racing,
Ilmenau Economics Discussion Papers, Vol. 25, No. 121, (2019).


\bibitem{Cheng}
Cheng, W., Rademaker, M., De Baets, B., H\"{u}llermeier, E.:
Predicting Partial Orders: Ranking with Abstention. In: Balcázar, J.L., Bonchi, F., Gionis, A., Sebag, M. (eds) Machine Learning and Knowledge Discovery in Databases. ECML PKDD 2010. Lecture Notes in Computer Science(), vol 6321. Springer, Berlin, Heidelberg.
\url{doi:org/10.1007/978-3-642-15880-3_20}



\bibitem{Conover}
Conover, W.J.:The rank transformation—an easy and intuitive way to connect many nonparametric methods to their parametric counterparts for seamless teaching introductory statistics courses. WIREs Comp Stat, 4: 432-438. (2012). \url{doi:10.1002/wics.1216}


\bibitem{Criado}
Criado, R., Garc\'{\i}a, E., Pedroche, F., Romance, M.:
A new method for comparing rankings through complex networks: Model and analysis of competitiveness of major European soccer leagues
Chaos: An Interdisciplinary Journal of Nonlinear Science.
23, 043114 (2013) \url{doi: 10.1063/1.4826446}


\bibitem{Diacon}
Diaconis, P., Graham, R.L.:
Spearman's Footrule as a Measure of Disarray.
J. R. Stat. Soc. B Met. 39, 262--268 (1977).
\url{doi:10.1111/j.2517-6161.1977.tb01624.x}


\bibitem{Fa06}
Fagin, R.,  Kumar, R., Mahdian, M., Sivakumar, D., Vee, E.:
Comparing Partial Rankings.
SIAM J. Discrete~Math. 20, 628--648 (2006).
\url{doi:10.1137/05063088X}


\bibitem{Garc}
Garcia-del-Barrio, P., Reade, J.J.:
Does certainty on the winner diminish the interest in sport competitions?
The case of formula one. Empir Econ 63, 1059–1079 (2022).
\url{https://doi.org/10.1007/s00181-021-02147-8}

\bibitem{Judde}
Judde, C., Booth, R., Brooks, R.:
Second Place Is First of the Losers: An Analysis of Competitive Balance in Formula One. Journal of Sports Economics, 14(4), 411–439 (2013)
\url{https://doi.org/10.1177/1527002513496009}


\bibitem{Kendall38}
Kendall, M.G.:
A New Measure of Rank Correlation. Biometrika 30, 81--89 (1938).
\url{doi:10.2307/2332226}

\bibitem{Kendall70}
Kendall, M.G.:
{\em Rank Correlation Methods}, 4th ed.; Griffin: London, UK, 1970.

\bibitem{KeBa39}
Kendall, M.G., Babington-Smith, B.:
The Problem of m Rankings.
Ann. Math. Stat. 10, 275--287 (1939). \url{doi:10.1214/aoms/1177732186}

\bibitem{KeHi}
Kendall, M. G., Hill, A. B.:
The Analysis of Economic Time-Series-Part I: Prices. Journal of the Royal Statistical Society. Series A (General), 116(1), 11–34.(1953)
\url{doi:10.2307/2980947}

\bibitem{Knuth}
Knuth, D. The art of computer programming. Vol 3. (2nd. Ed) 1988. Addison Wesley Longman

\bibitem{Kraus}
Krauskopf, T., Langen, M., B\"{u}nger, B.:
The search for optimal competitive balance in formula one,
CAWM Discussion Papers 38, University of Münster, Münster Center for Economic Policy (MEP). 2010.
\url{}


\bibitem{Manasis}
Manasis, V., Ntzoufras, I., Reade, J. J.:
Competitive balance measures and the uncertainty of outcome hypothesis in European football,
IMA Journal of Management Mathematics, (33), 1, 19--52 (2022)
\url{doi.org/10.1093/imaman/dpab027}


\bibitem{Mastro}
Mastromarco, C., Runkel, M.:
Rule changes and competitive balance in Formula One motor racing,
Applied Economics, 41(23), 3003–3014 (2009)
\url{doi: 10.1080/00036840701349182}

\bibitem{Owen}
Owen, P.D., Ryan, M., Weatherston, C.R.:
Measuring Competitive Balance in Professional Team Sports Using the Herfindahl-Hirschman
Index. Rev. Ind. Organ. 31, 289–302 (2007). \url{doi.org/10.1007/s11151-008-9157-0}



\bibitem{PeCrGaRoSa}
Pedroche, F., Criado, R., Garc\'{\i}a, E.,  Romance, M., S\'{a}nchez, V.E.
Comparing series of rankings with ties by using complex networks: An analysis of the Spanish stock market (IBEX-35 index).
Netw. Heterog. Media. 10, 101--125 (2015). \url{doi:10.3934/nhm.2015.10.101}

\bibitem{Pe20}
Pedroche, F.,  Conejero, J.A.
Corrected Evolutive Kendall's $\tau$ Coefficients for Incomplete Rankings with Ties: Application to Case of Spotify Lists.
Mathematics, 8(10), 1828, (2020).
\url{doi:10.3390/math8101828}

\bibitem{Shap}
Shapiro, S. S.,  Wilk, M. B. An Analysis of Variance Test for Normality (Complete Samples). Biometrika, 52(3/4), 591–611 (1965)
\url{doi.org/10.2307/2333709}

\bibitem{link-fia}
\url{https://www.fia.com}

\bibitem{season2012}
https://www.fia.com/events/fia-formula-one-world-championship/season-2012/classifications

%


\bibitem{Vecek}
Veček, N., Mernik, M., Črepinšek, M:
A chess rating system for evolutionary algorithms: A new method for the comparison and ranking of evolutionary algorithms,
Information Sciences, 277, 656-679 (2014). \url{doi:10.1016/j.ins.2014.02.154.}




\bibitem{Yoo}
Yoo, Y.; Escobedo, A.R.; Skolfield, J.K.:
A new correlation coefficient for comparing and aggregating non-strict and incomplete rankings.
Eur. J. Oper. Res. 285, 1025--1041 (2020). \url{doi.org/10.1016/j.ejor.2020.02.027}

\bibitem{Zimba}
Zimbalist, A. S.: Competitive Balance in Sports Leagues: An Introduction. Journal of Sports Economics,
3(2), 111–121. (2002). \url{doi:10.1177/152700250200300201}

\bibitem{wiki-esp} Primera División de España. Wikipedia. Retrieved 5/5/2023.
\url{https://es.wikipedia.org/wiki/Primera_División de España}


\bibitem{cero-premier} Premier League. Retrieved 5/5/2023.

\end{thebibliography}
\end{document}